%% file: main.tex
\documentclass[conference]{IEEEtran}
\IEEEoverridecommandlockouts
\usepackage{graphicx} 
\usepackage{amsmath}
\usepackage{xcolor}
\usepackage{algorithm}
\usepackage{algpseudocode}
\usepackage{multirow}
\usepackage{comment}

\algnewcommand\algorithmicforeach{\textbf{for each}}
\algdef{S}[FOR]{ForEach}[1]{\algorithmicforeach\ #1\ \algorithmicdo}

\algrenewcommand\algorithmicrequire{\textbf{Input:}}
\algrenewcommand\algorithmicensure{\textbf{Output:}}

\title{Low-Cost FlashAttention with Fused Exponential and Multiplication Hardware Operators}
\author{%
\IEEEauthorblockN{Kosmas Alexandridis, Vasileios Titopoulos, and Giorgos Dimitrakopoulos}
\IEEEauthorblockA{Integrated Circuits Lab 
Electrical and Computer Engineering
Democritus University of Thrace, Xanthi, Greece
\thanks{G. Dimitrakopoulos was supported by research grant of Siemens EDA to Democritus University of Thrace for ``HLS Research for Systems-on-Chip''.
K. Alexandridis was supported by an Industry Cooperation funded by Infineon Technologies Austria AG in the course of IPCEI Microelectronics}}}

\begin{document}

\maketitle

\begin{abstract}
Attention mechanisms, particularly within Transformer architectures and large language models (LLMs), have revolutionized sequence modeling in machine learning and artificial intelligence applications. To compute attention for increasingly long sequences, specialized accelerators have been proposed to execute key attention steps directly in hardware. Among the various recently proposed architectures, those based on variants of the FlashAttention algorithm, originally designed for GPUs, stand out due to their optimized computation, tiling capabilities, and reduced memory traffic. In this work, we focus on optimizing the kernel of floating-point-based FlashAttention using new hardware operators that fuse the computation of exponentials and vector multiplications, e.g., $e^x\, V$. The proposed ExpMul hardware operators significantly reduce the area and power costs of FlashAttention-based hardware accelerators. When implemented in a 28nm ASIC technology, they achieve improvements of 28.8\% in area and 17.6\% in power, on average, compared to state-of-the-art hardware architectures with separate exponentials and vector multiplications hardware operators.
\end{abstract}


\section{Introduction}
The Transformer architecture and its application in large language models (LLMs) have revolutionized machine learning, particularly in sequence modeling tasks like natural language processing~\cite{t5} and time-series analysis~\cite{tran_surv}. Their importance stems from the ability to selectively focus on relevant parts of the input through the attention mechanism, mimicking the human cognitive process of attending to specific information~\cite{base_attn}. This enables models to handle long-range dependencies effectively, a challenge that recurrent neural networks struggled with.

Roughly, attention achieves a dynamic, context-aware weighting of input features. Instead of treating all input elements equally, the attention mechanism calculates a score that represents the relevance of each element to the current processing step. This score is then used to weight the corresponding value, effectively amplifying the contribution of important elements and suppressing irrelevant ones. In essence, attention allows the model to focus to the most pertinent information, regardless of its position in the input sequence.

The efficiency of attention comes at a significant computational cost.
The quadratic complexity of the standard attention mechanism poses a significant bottleneck for processing long sequences in transformer models~\cite{longformer}, leading to prohibitive computational costs and limiting their application to shorter contexts. This limitation hinders the ability to capture crucial long-range dependencies necessary for tasks like document summarization and code generation. To address this, researchers have explored various methods to reduce computational burden, including approximating the full attention matrix through techniques such as sparse~\cite{sparse_attn}, linear~\cite{lin_attn}, and low-rank~\cite{low_rank_attn_2020} attention, which aim to balance accuracy and efficiency. 

Moreover, custom hardware accelerators~\cite{a3,lazy_softmax} have been developed to optimize key components of the baseline attention computation, such as matrix arithmetic~\cite{cosa} and softmax evaluation~\cite{softermax}, further enhancing processing speeds. These efforts collectively strive to make transformers more scalable and efficient for handling increasingly longer sequences.

In parallel, FlashAttention~\cite{fa,fa2,nsquared}, originally proposed for GPUs, addresses efficiently the computational bottlenecks inherent in the standard attention mechanism. It performs attention computation in tiles, avoiding the need to store the entire attention matrix. By recomputing attention scores on the fly and using optimized algorithms, flash attention reduces the memory footprint and improves computational efficiency.

In this work, we argue that leveraging FlashAttention in specialized hardware is essential for maximizing performance and efficiently processing extremely long sequences in real-world applications. To further enhance the hardware implementation of floating-point FlashAttention kernels, we design novel hardware operators that fuse exponential computation and vector multiplication. To support agile development, these operators are implemented using high-level synthesis (HLS) and made publicly available~\cite{our-repo}, enabling efficient design space exploration and faster verification.
The contributions of this work can be summarized as follows:
\begin{itemize}
\item 
The discrete steps of exponential function evaluation and the subsequent vector floating-point multiplication inherent in FlashAttention kernel are fused to one simplified ExpMul hardware operator. 
\item
Although the design of ExpMul includes a logarithmic quantization step, it does not hinder LLMs capabilities, as checked by running inference with Google's T5 model~\cite{t5} on the GLEU~\cite{glue} dataset.
\item
The experimental results highlight the efficiency of the proposed approach. Highly parallel FlashAttention accelerators implemented in a 28nm ASIC technology, achieve improvements of 28.8\% in area and 17.6\% in power, on average, compared to equivalent hardware architectures that implement exponential function evaluation and vector multiplications separately. 
\end{itemize}

\section{Computation of Attention and FlashAttention}
The attention mechanism facilitates the dynamic weighting of input sequence elements, enabling the model to capture long-range dependencies and contextual relationships. This process revolves around the interaction of query, key, and value vectors, derived from input embeddings~\cite{base_attn}.

\subsection{Attention Mechanism}

Input embeddings are linearly projected into three distinct spaces, each representing a different aspect of the attention mechanism. These projections produce the query ($Q$), key ($K$), and value ($V$) matrices, which are then used to compute the attention weights and output. In self-attention, these matrices are derived from the same input sequence, allowing the model to assess relationships between different positions within the sequence itself.
In summary, attention is computed as follows:
\begin{equation}
\text{Attn}(Q, K, V) = \text{softmax}(Q K^T/\sqrt{d}) V
\label{e:basic-attn}
\end{equation}
Similarity scores $Q K^T$ are scaled by the square root of the key vector dimension $d$ to mitigate the issue of vanishing gradients that can arise, when the scores become excessively large. For the rest of the paper, to simplify notation, we leave out this constant scaling. 
To detect the most relevant of tokens in~\eqref{e:basic-attn} the softmax function is applied to the similarity scores for the same query. The final output is produced by multiplying the attention score matrix derived for all queries to the matrix of value vectors, where each value’s contribution to the output is heavily based on the corresponding scores of the attention score matrix. 

In practice, the attention mechanism is applied across multiple heads in parallel~\cite{base_attn}, allowing the model to comprehend complex relationships, make it more robust to erroneous generations and enable its parallel execution across heads, thus enhancing performance.

The computation of attention employing the lazy softmax division concept~\cite{lazy_softmax, nsquared} is shown in Alg.~\ref{alg:attn-lazy}. The first part computes the dot product of a query vector with all key vectors. Also the maximum score is identified that is needed for computing a numerically-stable softmax normalization. Exponentiating each score, as needed by softmax, may lead to infinite values that would propagate until the output and ruin the final result. To avoid this numerical instability, softmax is implemented by subtracting the maximum from all scores. This effectively removes the problem of overflow and numeric instability, while preserving the fundamental properties of the softmax function.

\begin{algorithm}[t]
\caption{Attention with lazy softmax division}
\label{alg:attn-lazy}
\begin{algorithmic}[1]
\ForEach {query $\vec{q}$}
\For{$i = 1:N$} 
\State $s_i \gets \text{dot}(\vec{q}, \vec{k}_i)$
\State $m_i \gets \max(m_{i-1}, s_i)$
\EndFor
\State $\ell_0 \gets 0$
\For{$i = 1:N$} 
\State $\ell_i \gets \ell_{i-1} +e^{s_i - m_N}$
\State $\vec{o}_i \gets \vec{o}_{i-1} + e^{s_i - m_N}\cdot \vec{v}_i$
\EndFor
\State $\text{attn}(\vec{q}, K, V) \gets \vec{o}_N/ \ell_N$
\EndFor
\end{algorithmic}
\end{algorithm}

In the following, the output vector $\vec{o}_i$ is computed incrementally as the weighted sum of each value vector and the corresponding exponential score decremented by the maximum score. In parallel, the sum of all exponentials $\ell_i$ is accumulated. The final attention vector for one query vector $\vec q$ is computed by dividing the final output vector by the corresponding sum-of-exponents $\ell_N$.

\subsection{Attention Hardware Accelerators State-of-the-Art}
The traditional attention mechanism faces three primary performance bottlenecks: high memory bandwidth demands due to the retrieval of query, key, and value matrices, particularly with long sequences; substantial computational overhead from the softmax operation across the full sequence; and limited parallelism due to reduction operations within softmax, such as maximum value determination and exponential summation.

To mitigate memory traffic, attention accelerators employ a strategy of keeping key and value vectors static in local SRAM, while streaming query vectors to calculate attention scores~\cite{a3,keller,lu}. This approach minimizes memory fetches by only reloading query vectors for each key and value vector. However, this method's effectiveness diminishes with increasing sequence lengths, as key, value, and score matrices are replaced into slower DRAM, leading to performance degradation. 
This issue is particularly relevant in modern NLP tasks that often involve extensive sequences~\cite{longformer}. To reduce the memory access overhead other approaches focus on in-memory computation of attention~\cite{x-former}.

To decouple computational resources and local memory from the sequence length, several designs~\cite{lazy_softmax,cosa} accumulate partial softmax results for each column of the attention scores. 
This method maintains accuracy while reducing the need to buffer and compute softmax results for the entire sequence at once, avoiding memory spills. 

Further optimization techniques involve computation skipping based on token similarity to reduce latency and power~\cite{elsa,tsacc}, as well as quantization to minimize the cost of data transfers and enhance accelerator efficiency~\cite{swiftron}.

\section{FlashAttention-based Accelerators}
FlashAttention~\cite{fa}, inspired by online softmax computation~\cite{online-softmax}, restructures attention computation into an online process. Alg.~\ref{alg:flash-attn2} illustrates its more efficient variant, FlashAttention-2, which further optimizes performance by postponing the softmax division until the end, using the corresponding sum of exponents~\cite{lazy_softmax, nsquared}. The key distinction from baseline attention (Algorithm~\ref{alg:attn-lazy}) is that FlashAttention-2 computes all necessary variables online within the same inner loop, eliminating the need to first determine the maximum of all attention scores. For large sequence lengths ($N$), this online computation of softmax components is crucial for maintaining performance efficiency.

\begin{algorithm}
\caption{FlashAttention-2 with delayed softmax division}\label{alg:flash-attn2}
\begin{algorithmic}[1]
\ForEach {query $\vec{q}$}
\For{$i = 1:N$} 
\State $s_i \gets \text{dot}(\vec{q}, \vec{k}_i)$
\State $m_i \gets \max(m_{i-1}, s_i)$
\State $\ell_i \gets \ell_{i-1}e^{m_{i-1}-m_i}+e^{s_i-m_i}$
\State $\vec{o}_i \gets \vec{o}_{i-1} e^{m_{i-1}-m_i}+\vec{v}_i e^{s_i-m_i}$
\EndFor
\State $\text{attn}(\vec{q}, K, V) \gets \vec{o}_N/\ell_N$
\EndFor
\end{algorithmic}
\end{algorithm}

During each iteration, the dot product of the query vector and a key vector yields a similarity score, denoted as $s_i$. Subsequently, $m_i$ is determined as the current maximum attention score. Then, $\ell_i$ incrementally accumulates the sum of the exponentials of each $s_i$ decremented by the present maximum value. The multiplication by $e^{m_{i-1}-m_i}$ in the calculation of $\ell_i$ effectively adjusts the prior maximum value whenever the current maximum $m_i$ is larger than the previous maximum $m_{i-1}$. Similarly, the output vector $\vec{o}_i$ is updated by adding the new value vector $\vec{v}_i$ weighted by its softmax importance, to the maximum-adjusted preceding output vector, $\vec{o}_{i-1} e^{m_{i-1}-m_i}$.At the end, to finalize attention computation for this query vector the output $\vec{o}_N$ is divided by the total sum of exponents accumulated in parallel in $\ell_N$.

\begin{figure}
\centering
\includegraphics[width=0.98\columnwidth]{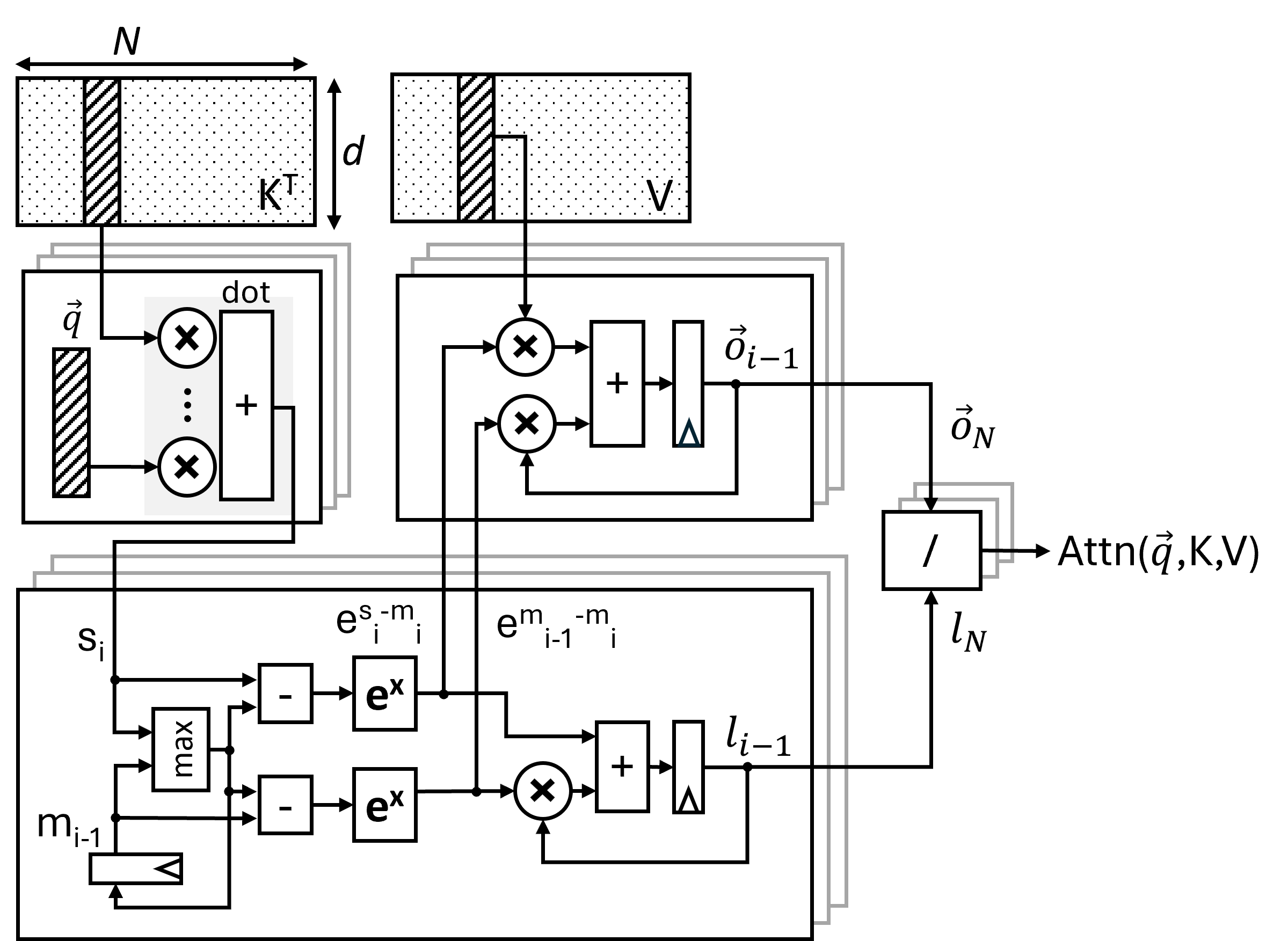}
\caption{A block-parallel hardware architecture for FlashAttention-2 kernel.}
\label{f:flashattn2-hw}
\end{figure}

The FlashAttention-2 kernel shown in Alg.~\ref{alg:flash-attn2} involves two for loops that can be unrolled to enhance parallelism and computational throughput. 
To avoid any serial dependencies, we unroll the outer loop, allowing the FlashAttention-2 kernel to process multiple query vectors in parallel within the same blocks of key and value vectors.
Fig.~\ref{f:flashattn2-hw} shows the derived parallel hardware structure 
for FlashAttention-2. A block of query vectors is pre-loaded locally, while key vectors are provided sequentially to all blocks to compute the corresponding dot products~\cite{ol-align}. 
This process leads to the calculation of a new maximum value and an updated running sum of exponents. The output vectors for different query vectors are updated in parallel as value vectors are streamed into the accelerator. After processing all key and value vectors, the attention for each query vector is computed through a final division operation. The computation finishes once all query vectors have been processed.
For FlashAttention-2 kernels, we assume that we can read from local memories in each cycle one key and one value vector of $d$ elements each that corresponds to the size of the hidden dimension of attention.

FlashAttention-2 does not require a unified softmax hardware unit, since exponentiations and the final division are computed separately. Many approaches have been used for computing those two non-linear operations in hardware. Variants include piece-wise linear approximations after appropriate range reduction~\cite{koca}, transformations based on logarithmic quantization~\cite{sole} or other approximations~\cite{peano-vit}.

\section{Fused Exponential and Vector Multiplication Hardware Operators}
Our goal is to enhance the last part of the FlashAttention-2 kernel (line 6 in Alg.~\ref{alg:flash-attn2}), which incrementally updates the output by performing two vector multiplications and one vector addition. To achieve this, we plan to fuse the exponential calculations with the vector multiplications, utilizing new, low-cost hardware operators.

\subsection{Merging sum-of-exponents and output update}
The first step in the planned simplification of the FlashAttention-2 kernel is to merge the incremental computation of the sum-of-exponents (performed in line 5 of Alg.\ref{alg:flash-attn2}) with the incremental computation of the output vector (line 6 in Alg.\ref{alg:flash-attn2}). Both updates involve the same operations and can be written in a vector-merged form as follows:
\begin{equation}
\begin{bmatrix}
\ell_i \\           
\vec{o}_i
\end{bmatrix}=
\begin{bmatrix}
\ell_{i-1}\cdot e^{m_{i-1}-m_i}+1\cdot e^{s_i-m_i} \\           
\vec{o}_{i-1}\cdot e^{m_{i-1}-m_i}+\vec{v}_i\cdot e^{s_i-m_i}
\end{bmatrix}
\label{e:merged}
\end{equation}
Increasing by one element the output $\vec{o}_i$ and the value vector $\vec{v}_i$, i.e., $o^*_i = [\ell_i\quad \vec{o}_i]$ and $v^*_i=[1\quad \vec{v}_i]$ the merged incremental update of the sum-of-exponents and the output shown in~\eqref{e:merged} can be written as:
\begin{equation}
o^*_i = o^*_{i-1}\cdot e^{m_{i-1}-m_i}+v^*_i\cdot e^{s_i-m_i}
\label{e:merged-output}
\end{equation}

The unified form of~\eqref{e:merged-output} demonstrates that the incremental computation of the output involves two scalar × vector multiplication operations, where the scalar value is determined after evaluating the exponential function. To simplify the hardware for this computation, we propose fusing the separate exponential and vector multiplication operations into a single unified hardware operator:
\begin{equation} 
\text{ExpMul}(x, V) = e^x\, V
\label{e:exp-mul}
\end{equation}
This would allow us to rewrite~\eqref{e:merged-output} as follows:
\begin{equation}
o^*_i = \text{ExpMul}(m_{i-1}-m_i, o^*_{i-1})+\text{ExpMul}(s_i-m_i, v^*_i)
\label{e:output-exp-mul}
\end{equation}

\input{fused_expv}

\section{Evaluation}
Experimental evaluation aims to examine the impact of the new fused ExpMul hardware operators on real machine-learning applications and also highlight the achieved hardware savings.

\subsection{Verification of LLM accuracy}
Since ExpMul involves logarithmic quantization, it introduces a numerical approximation. To verify that this approximation does not hinder the capabilities of a Large Language Model (LLM), we run ten representative NLP benchmarks from the GLUE~\cite{glue} dataset, using Google's Flan-T5 LLM.
For comparison, we considered four implementations: (a) Models
’FP32’ and ’BF16' use the default attention mechanism of T5 and compute all operations including separate multiplications and function evaluation using single-precision and BFloat16~\cite{bfloat16} floating-point arithmetic; (b) Models ’FP32-ExpMul’ and 'BF16-ExpMul' employ floating-point arithmetic for all parts of the model except for the proposed ExpMul operator, where computations are done in 16-bit integer arithmetic, as discussed in Section~\ref{s:fused_op_expl}.
To run inference with the LLM we utilized Microsoft's PromptBench workflow~\cite{promptbench}.

Table~\ref{t:accuracy} presents the accuracy results for all four examined variants. All models exhibit comparable performance in terms of inference accuracy and the F1-score metric. In some cases, reduced-precision models (‘{FP32, BF16}-ExpMul’) outperform their full-precision counterparts. This phenomenon is a known artifact of lower-precision models~\cite{i-bert}. While it does not indicate that the model is inherently superior, it confirms that the approximation used to formulate the ExpMul operators does not degrade performance.

\subsection{Hardware Savings}
To evaluate the hardware savings of the proposed ExpMul hardware operators we implemented two variants of the FlashAttention-2 kernel. The two designs correspond to the main block shown in the foreground of Fig.~\ref{f:flashattn2-hw} and Fig.~\ref{f:ExpMul_hw}, respectively. The implementation involves 
various hidden dimension sizes ($d$) and two floating-point formats: FP32 and BFloat16. The baseline FlashAttention kernel performs exponent and multiplication operations separately. Exponent function evaluation is implemented with piece-wise linear approximation in the reduced input range inherent to attention and used also for the ExpMul operators.

In practice, the total cost of the unrolled hardware serving multiple query vectors in parallel will be the cost for one query multiplied by the number of parallel query vectors served. The query vectors are preloaded separately in the architecture, while the key and value vectors are loaded and broadcasted to all parallel blocks. 

Both hardware blocks were implemented in C++
(publicly available at~\cite{our-repo})
and synthesized into Verilog using Catapult HLS with a 28-nm standard-cell library.
Both designs operate at the same pipelined latency with a clock frequency of 500 MHz. 
Increasing the hidden dimension of each attention accelerator increases also the latency of computation. For the examined sizes of the hidden dimension, for one key and value vectors pipelined latency with initiation interval of 1, ranges between 8 and 12 cycles. Next, Verilog was synthesized using the Cadence digital implementation flow, while power consumption was estimated with the PowerPro power analysis and optimization tool. The reported power consumption represents the average power measured after executing attention kernels for the T5 Large Language Model and GLUE benchmarks utilizing the PromptBench workflow.
\begin{figure}[t]
\centering
\includegraphics[width=0.92\columnwidth]{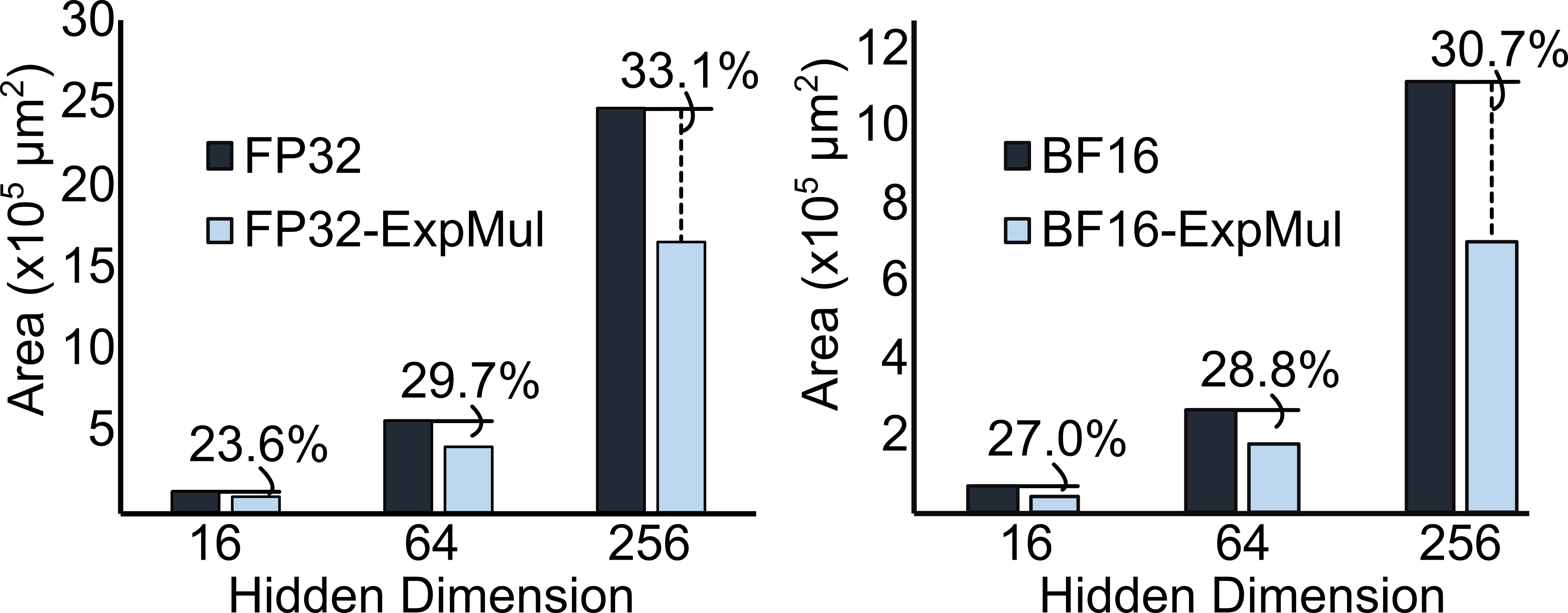}
\caption{The hardware area at 28 nm for the FlashAttention-2 kernel, evaluated across various hidden dimension sizes, with and without ExpMul operators, for computing attention on a single query using FP32 and BFloat16 floating-point formats.}
\label{f:area}
\end{figure}

\begin{figure}[t]
\centering
\includegraphics[width=0.92\columnwidth]{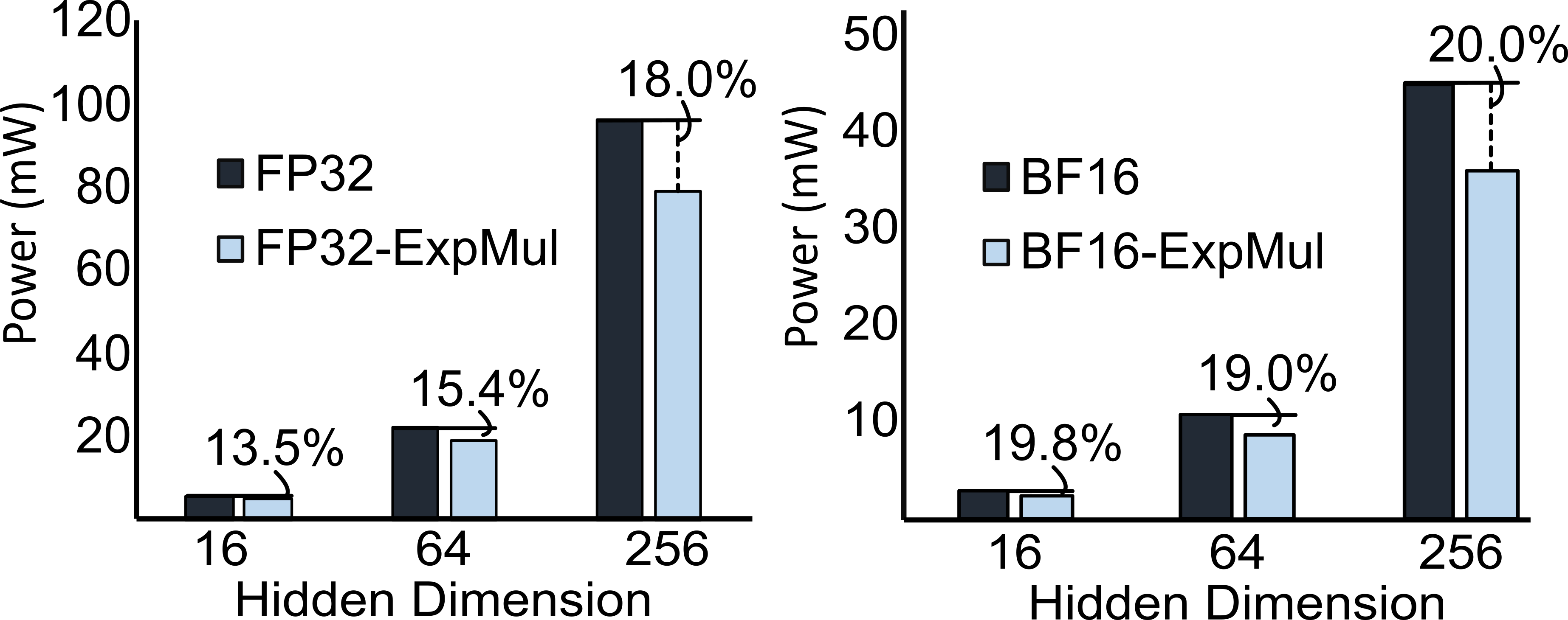}
\caption{The average power consumption of the FlashAttention-2 kernel, evaluated across various hidden dimension sizes, with and without ExpMul operators, for computing attention on a single query using FP32 and BFloat16 floating-point formats. Memory and I/O power are excluded, as they are identical in both designs.}
\label{f:power}
\end{figure}

Figs.~\ref{f:area} and~\ref{f:power} show the area and power of the FlashAttention-2 kernel with and without ExpMul hardware operators, for the two examined floating-point data type and for three hidden dimension sizes $d=\{16,\ 64,\ 256\}$.
Power estimation excludes memory power and focuses solely on the average power consumption of the computation kernel. The memory power in both cases is expected to be identical, as both approaches implement the same FlashAttention-2 algorithm using the same computation order and data flows. The difference lies solely on how the computation kernel is executed internally.

As shown in Fig.~\ref{f:area}, utilizing ExpMul operators reduces the hardware area by more than 28.8\% on average in all examined cases. 
These savings come from the simplified computation kernel that does not require area expensive floating-point multipliers and or additional exponential function evaluations.
In the output update module, $d$ multiplication and exponentials are converted to ExpMul operators, while reforming the floating-point output requires only $d$ exponent subtractions.

Simplifying hardware architecture improves energy efficiency, reducing power consumption by over 17.8\% on average. As the hidden dimension $d$ increases, power savings increase due to replacing power-hungry floating-point multipliers with simpler ExpMul operators that use integer arithmetic.

\section{Conclusions}
FlashAttention kernel simplifies attention computation by fusing the softmax function evaluation and output computation in an online manner.
This transformation, originally proposed for GPUs, allows attention computation in tiles with sizes independent of the sequence length.
In this work, our goal is to leverage the FlashAttention-2 algorithm, an optimized version of the baseline FlashAttention, to design hardware accelerators that not only benefit from all the advantages of FlashAttention but also save significant area and power. This is achieved by employing the fused exponential and vector multiplication operators, which lead to efficient hardware architectures without compromising the accuracy of state-of-the-art LLM applications.

\bibliographystyle{IEEEtran}
\bibliography{ref}

\end{document}

%% file: fused_expv.tex
\subsection{Fused Exponential-Vector Multiplication Operator}
\label{s:fused_op_expl}
In this work, we focus on the floating-point arithmetic implementation of FlashAttention. To efficiently implement the newly introduced ExpMul operator, we utilize logarithmic quantization techniques originally proposed for softmax hardware implementations~\cite{sole}. These techniques allow us to replace the computationally expensive floating-point exponential and multiplication operations required by the ExpMul operator with hardware-friendly integer shift-and-add operations.

The ExpMul operators in FlashAttention-2 receive a difference of scalar values that is always less or equal to 0 and a vector of floating-point data. Since the value of $x$ in~\eqref{e:exp-mul} belongs to $(-\infty, 0]$, it effectively limits $e^x$ in the range of $(0,1]$. This characteristic
fits logarithmic quantization, which applies the $\log_2$ function and returns as a result, its rounded negative value.
\begin{align}
\text{Log2Exp}(X) = \lfloor-\log_2 e^X \rceil=\lfloor-X\cdot\log_2 e\rceil
    \label{e:log2exp}
\end{align}
Since $X$ is a floating-point number, the multiplication with the constant $\log_2 e$ should be performed using floating-point arithmetic. 

To simplify the computation, we convert the value of $X$ to its fixed-point equivalent $\hat{X}$. In general, converting a floating-point number to its fixed-point representation requires many integer bits to accommodate the wide range of floating-point arithmetic~\cite{wide-acc}. However, in this case, we can safely clip 
$X$ to a much smaller dynamic range, as it will only be used for the computation of 
$e^X$ for negative values of $X$. In this range, $e^X$ quickly converges to 0, even for small negative 
$X$ values, i.e., $e^{-15}=3.1\cdot 10^{-7}$.
In other words we can convert $X$ to a fixed-point number with a small number of bits without the need to consider its full dynamic range. Thus, instead of fully quantizing $X$ we simply constrain its value before logarithmic quantization.
\begin{equation}
\hat{X} = \text{Fixed}(\text{Clip}(X,-15, 0)) \nonumber
\end{equation}
Converting to a fixed-point number the clipped value of $X$ requires a small number of bits.
For this conversion, we use a 16-bit fixed-point number to keep the corresponding hardware overhead small. Since $X$ is going to be multiplied by a constant value its range after clipping will change to $[-21.64,\ 0]$. To account for this new range we assign 6 bits for the integer part and the rest for the fraction part. 

After the clipping and quantization of $X$, Eq.~\eqref{e:log2exp} becomes
\begin{align}
    \text{Log2Exp}(X)=\lfloor-\hat{X}\cdot\log_2 e\rceil \nonumber
\end{align}
Since $\hat{X}$ is a fixed-point number the multiplication with the constant value $\log_2 e$ can be approximated by integer shift--and--add operations~\cite{sole,koca}. 
Following the approach presented in~\cite{sole}, we get that:
\begin{align}
    \text{Log2Exp}(X)=-\lfloor\hat{X}+\hat{X}\gg 1-\hat{X}\gg4\rceil
    \label{e:log2exp_fx}
\end{align}

With the help of~\eqref{e:log2exp_fx} the ExpMul operator of~\eqref{e:exp-mul} can be rewritten as:
\begin{align}
    \text{ExpMul}(x,V)=2^{-\hat{L}}\, V, \text{where  } \hat{L}= \text{Log2Exp}(x).
    \label{e:exp-mul-fx}
\end{align}

\noindent Multiplying the floating-point number $V=(-1)^{S_V} 2^{E_V-b} (1+M_V)$, where $S_V, E_V, M_V$ represent its sign, exponent and mantissa field, respectively, with $2^{-\hat{L}}$ in~\eqref{e:exp-mul-fx} is equivalent to subtracting $\hat{L}$ from the exponent of $V$. In case of an overflow the resulting FP number is set to 0.
More specifically,
\begin{align}
\text{ExpMul}(x,V)
&=2^{-\hat{L}}\,V = 2^{-\hat{L}}\, (-1)^{S_V}\, 2^{E_V-b}\, (1+M_V)  \nonumber \\
&=(-1)^{S_V}\,2^{E_A-b-\hat{L}}\,(1+M_V)
\nonumber \\
&=(-1)^{S_V}\,2^{E_A-b-\text{Log2Exp(x)}}\,(1+M_V)
\label{e:exp-mul-fin}
\end{align}
The result computes both $e^x\, V$ and at the same time returns the result directly in floating-point format without requiring any addition conversion (or dequantization) step.

Alg.~\ref{alg:exp-mul} summarizes the computational steps that are performed by the ExpMul operator to calculate $e^x\, V$ and Alg.~\ref{alg:flash-attn2-expmul} using~\eqref{e:merged-output},~\eqref{e:output-exp-mul}, and~\eqref{e:exp-mul-fin} illustrates how the new hardware operator is integrated in FlashAttention-2 kernel.
\begin{algorithm}
\caption{ExpMul(x, V)}
\label{alg:exp-mul}
\begin{algorithmic}[1]
\Require{$x < 0$ and a vector $V$ of $N$ elements}
\Ensure{A vector $Out$ where $Out[i]=e^x V[i]$}
\For{$i\gets 1:N$}
\State $S_V,\ E_V,\  M_V\  \gets \text{extract}(V[i])$
\State $\hat{x} \gets \text{Fixed}(\text{Clip}(x, -15, 0))$
\State $\hat{L} \gets -\lfloor\hat{x}+\hat{x}\gg 1-\hat{x}\gg4\rceil$
\State $Out[i] \gets \text{Float}(S_V,\ E_V-\hat{L},\ M_V)$
\EndFor
\end{algorithmic}
\end{algorithm}

\begin{algorithm}
\caption{FlashAttention2 with Fused ExpMul Operators}
\label{alg:flash-attn2-expmul}
\begin{algorithmic}[1]
\ForEach {query $\vec{q}$}
\For{$i = 1:N$} 
\State $s_i \gets \text{dot}(\vec{q}, \vec{k}_i)$
\State $m_i \gets \max(m_{i-1}, s_i)$
\State $o^*_i \gets 
        \begin{aligned}[t]
           & \text{ExpMul}(m_{i-1}-m_i, o^*_{i-1})\, + \\
           & \text{ExpMul}(s_i-m_i, v^*_i)
        \end{aligned}$
\EndFor
\State $[\ell_N \quad \vec{o}_N] \gets o^*_N$
\State $\text{attn}(\vec{q}, K, V) \gets \vec{o}_N/\ell_N$
\EndFor
\end{algorithmic}
\end{algorithm}

The proposed ExpMul operators not only remove expensive floating-point operations like exponent function evaluation and multiplication but
perform this step with low-cost quantization and without any dequantization step.
Traditional quantization approaches~\cite{swiftron} would utilize additional quantization logic to transfer operation to and from the integer domain, thus paying the extra hardware cost of this transformation. 
Also, even if operating in the integer domain, the computation would still involve costly multiplication and exponential operations.
The proposed approach removes both overheads and simplifies significantly the computation in FlashAttention-2.

\begin{figure}[h!]
\centering
\includegraphics[width=0.8\columnwidth]{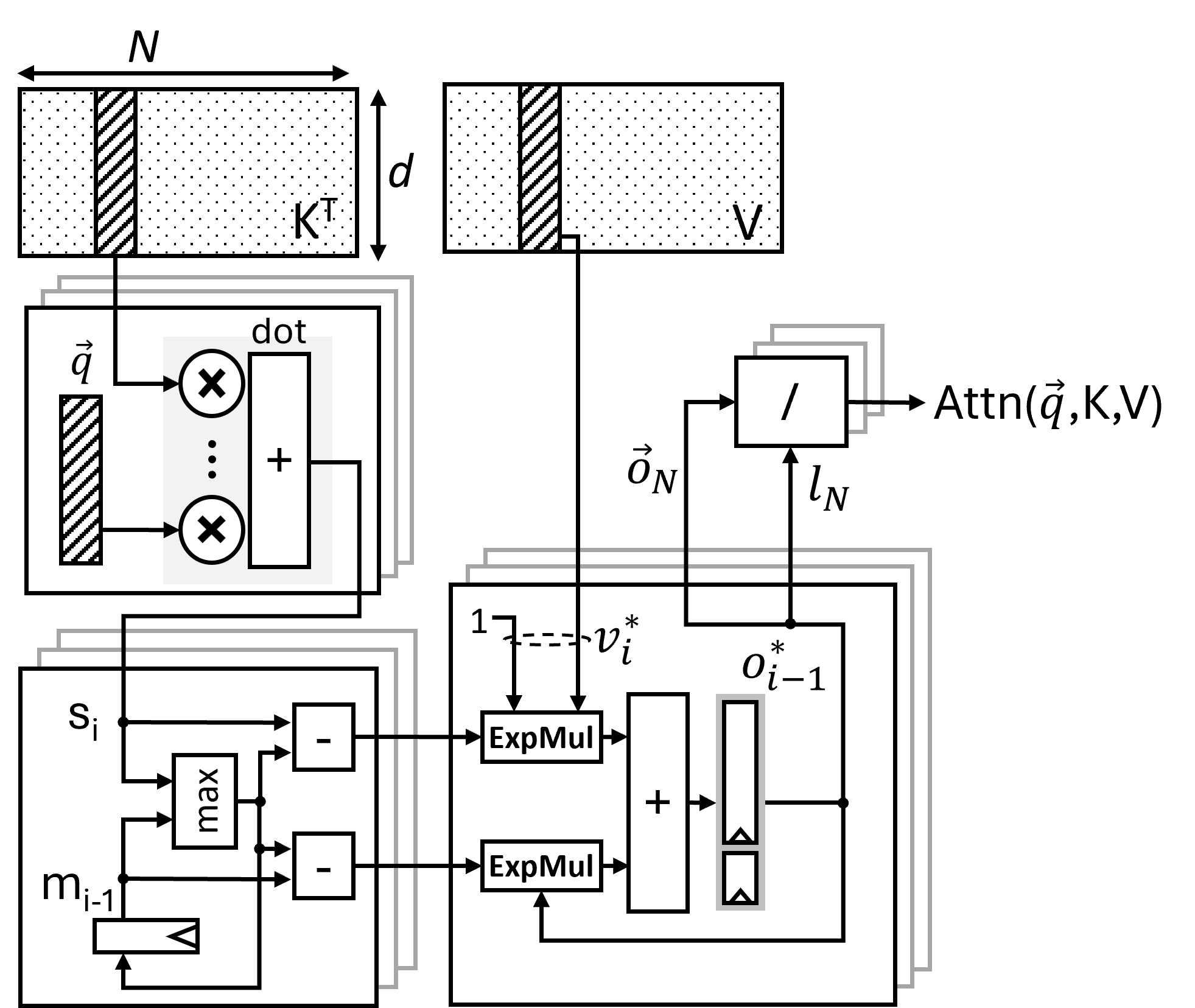}
\caption{The FlashAttention-2 kernel optimized with the proposed ExpMul operators. According to Alg.~\ref{alg:exp-mul} each ExpMul operator involves clipping, integer shift-and-add operations and an exponent increment producing directly floating-point results without further dequantization.}
\label{f:ExpMul_hw}
\end{figure}

\begin{table*}[t!]
    \centering
    \caption{Performance of Google's FLAN-T5 LLM model for 10 benchmarks of the GLUE dataset~\cite{glue}. }
    \begin{tabular}{|c|l||c|c|c|c|c|c|c|c|c|c|}
        \hline
          \multicolumn{2}{|c||}{Benchmarks} &{\bf STS-2}&{\bf MNLI-m}& {\bf MNLI-mm}&{\bf QQP}&{\bf QNLI}&{\bf CoLA}&{\bf MRPC}& {\bf RTE}& {\bf WNLI}& {\bf STS-B}  \\\hline \hline
        \multirow{4}{*}{Accuracy (\%)}
        &FP32        & 92.1 & 87.5   & 84.2  &93.1 & 93.3 & 72.0 & 86.0 & 74.3 & 62.0 & 92.0 \\
        &FP32-ExpMul & 92.1 & 87.5   & 84.2  &93.1 & 93.3 & 72.0 & 86.0 & 74.3 & 62.8 & 92.0 \\
        &BF16        & 91.1 & 87.3   & 83.3  &93.1 & 93.3 & 72.0 & 84.0 & 73.8 & 62.0 & 91.0 \\
        &BF16-ExpMul & 91.2 & 87.3   & 82.1  &93.1 & 93.3 & 69.0 & 88.0 & 73.8 & 62.0 & 90.0 \\
        \hline \hline
        \multirow{4}{*}{F1-score}
        &FP32        &0.921 & 0.794  & 0.845  &0.930 & 0.933 & 0.830 & 0.900 & 0.726 & 0.690 & - \\
        &FP32-ExpMul &0.931 & 0.800  & 0.833  &0.930 & 0.920 & 0.840 & 0.900 & 0.738 & 0.710 & - \\
        &BF16        &0.910 & 0.780  & 0.845  &0.930 & 0.900 & 0.800 & 0.890 & 0.725 & 0.690 & - \\
        &BF16-ExpMul &0.910 & 0.780  & 0.833  &0.930 & 0.900 & 0.780 & 0.920 & 0.733 & 0.690 & - \\
        \hline
    \end{tabular}
    \label{t:accuracy}
\end{table*}

The parallel hardware organization of FlashAttention-2 kernel employing the proposed ExpMul operators is shown in Fig.~\ref{f:ExpMul_hw}. The organization of the new hardware unit is the same as the original FlashAttention-2 accelerator shown in Fig.~\ref{f:flashattn2-hw}. However, we can highlight one key difference. The exponent function hardware operators and floating-point multipliers are replaced by a single ExpMul unit which performs those two operations in a fused manner using integer only arithmetic and preparing the result in floating-point as needed by subsequent operations.